\documentclass[aps,prl,superscriptaddress,twocolumn,amsfonts,amsmath,a4paper,showpacs]{revtex4}
\usepackage{graphicx}
\usepackage{multirow}
\usepackage{booktabs}

\begin{document}


\title{Tailoring the Phonon Band Structure in Binary Colloidal
  Mixtures}

\author{Julia Fornleitner}
\affiliation{Center for Computational Materials Science and Institut
f\"ur Theoretische Physik, Technische Universit\"at Wien, Wiedner
Hauptstra{\ss}e 8-10, A-1040 Vienna, Austria}
\affiliation{IFF Theorie II, Forschungszentrum J{\"u}lich, 
D-52425 J{\"u}lich, Germany}

\author{Gerhard Kahl}
\affiliation{Center for Computational Materials Science and Institut
f\"ur Theoretische Physik, Technische Universit\"at Wien, Wiedner
Hauptstra{\ss}e 8-10, A-1040 Vienna, Austria}

\author{Christos~N.~Likos}
\affiliation{Institute of Theoretical Physics,
Heinrich-Heine University of D\"{u}sseldorf,
D-40225 D{\"u}sseldorf, Germany}
\affiliation{Faculty of Physics, University of Vienna,
Sensengasse 8/12, A-1090 Vienna, Austria}

\begin{abstract}
We analyze the phonon spectra of periodic structures formed by 
two-dimensional mixtures of dipolar colloidal particles.
These mixtures
display an enormous variety of complex ordered configurations
[J.~Fornleitner {\it et al.}, Soft Matter {\bf 4}, 480 (2008)],
allowing for the systematic investigation of the ensuing phonon
spectra and the control of phononic gaps. We show how the shape
of the phonon bands and the number and width of the phonon gaps
can be controlled by changing the susceptibility ratio, the
concentration and the mass ratio between the two components. 
\end{abstract}

\date{\today}

\maketitle

Materials with a band gap in their spectrum of transmitted sound waves have
been the focus of intensive research recently. 
Most often, these ``phononic crystals'', termed in
analogy to the more familiar photonic crystals, are analyzed and fabricated
on the basis of macroscopic approaches: following methods 
from scattering theory, 
materials with a periodic modulation in their elastic properties and/or
density are assembled, so that mismatches in the speed of sound and
destructive interference lead to the desired gaps \cite{Pen02,Che06,Sti08}.
An alternative and very promising route to understand and engineer 
phononic crystals is to focus on the microscopic properties that directly shape
the dispersion curves, i.e., the interparticle interactions.
The latter uniquely determine {\it both} the ground-state configuration of the
system {\it and} its elemantary excitations above the same, i.e., the phonons.
A class of materials where this microscopic approach is particularly 
fruitful are
colloidal crystals. In such systems, the interactions are 
tunable and versatile.
It has been shown that the dispersion curves of two-dimensional colloidal
crystals can be shaped and controlled by suitable external substrate potentials
\cite{Gru07,Bau07,Bau08}. 
Here, we focus on systems that exhibit phononic gaps without
the presence of external fields, namely {\it mixtures}: 
the non-trivial unit cells of
their ground-states give rise to optical branches in the phonon band structure
(PBS), thus opening the possibility to induce gaps by suitable adjustments in
the interactions. Here, we explore the possibilities
to tune the PBS's of a 
binary mixture of dipolar colloids via
changes in the susceptibility ratio and concentration.

Binary dipolar monolayers are readily available to experiments
\cite{Zah99,Zah00,Zah03,Kei04,Hof06Ebe,Ebe08} and exhibit a rich wealth of
stable ordered structures \cite{For08LoV,For09}. Experimental realizations
employ super-paramagnetic particles of different 
susceptibilities $\chi$, which are trapped at the interface of a pendant water
droplet to ensure a planar geometry \cite{Ebe08}. An external magnetic field
$\mathbf{B}$ applied perpendicular to the water-air-interface induces magnetic
moments in the colloidal particles parallel to the external field,
$\mathbf{M}_i=\chi_i\mathbf{B}$, with $i=A,B$ labeling the particle
species. We emphasize that here the external field serves only as a means
to influence the {\it two-body interactions} in the Hamiltionian and does
not act on the systems as a {\it one-body potential}, as is the case of
laser beams in Refs.~\cite{Gru07,Bau07,Bau08}. For the sake of simplicity, we assume that
the physical size of the two particle species is the same and given by
their common diameter $\sigma$, the disparity in their interactions arising
from different degrees of doping with ferromagnetic material, such as
${\rm Fe}_2{\rm O}_3$.
Setting
$m_i=\chi_i/\chi_A \leq 1$, the ideal dipole-dipole repulsion acting in the
mixture can be written as $\Phi_{ij}(x)=\varepsilon{m_im_j}/{x^3}$, $i,j=A,B$,
with $x=r/\sigma$ and
$\varepsilon=\mu_0\chi_A^2 |{\bf B}|^2/(4\pi\sigma^3)$. The
ground state is thus determined by the asymmetry in
dipole strength, given by $m=\chi_B/\chi_A < 1$, and by the composition of
the mixture, 
$C=n_B/(n_A+n_B)$, with $n_{A(B)}$ being the number of strong
(weak) dipoles per unit cell.

Colloids in solution obey Langevin dynamics and their full
equations of motion include the interparticle forces, the random collisions
with the solvent and hydrodynamic interactions~\cite{hur82}, involving 
the masses of the particles, the frictions constants and the solvent properties {\it via}
the Navier-Stokes equations. Accordingly, the full dynamics of colloidal
crystals have been thoroughly analyzed by theory and experiment in the
past~\cite{hur82,der92,ohs01,pia91}. Here we focus only on the excitation
spectrum of the crystal, which is expressed in terms of the dispersion
relation $\lambda({\bf q})$, where ${\bf q}$ denotes a wavevector in the
first Brillouin zone and $\lambda({\bf q})$ the eigenvalue. Although
oscillations in a colloidal crystal are overdamped, the quantities
$\lambda({\bf q})$ are still experimentally measurable by means of the
equipartition theorem~\cite{Gru07}, a strategy that has been successfully applied
to the one-component version of the system at hand~\cite{Kei04}.

Accordingly, we apply harmonic lattice theory~\cite{Ash76,Mar00}
and obtain the dispersion curves
$\lambda(\mathbf{q})$ by solving the eigenvalue equation
$\lambda({\bf q})c_{\nu i}=\sum_{\nu'i'}
\tilde{D}_{\nu i,\nu' i'}(\mathbf{q})c_{\nu' i'}$,
where $\tilde{D}_{\nu i,\nu' i'}(\mathbf{q})$ is the Fourier transform of the
dynamical matrix $D_{\nu i,\nu' i'}(n,n')$, defined as:
\begin{equation}\label{eqn_DynMatrix}
D_{\nu i,\nu'i'}(n,n')=\frac{\partial^2\Phi(\mathbf{r}^N)}{\partial u_{n\nu
    i}\partial u_{n'\nu'i'}}|_{\mathbf{u}^N=0}.
\end{equation}
Here, the index $n$ is used to label the unit cell in the lattice and $\nu$ runs
over all particles within one unit cell, $\nu=1,...,n_A+n_B$,
$\mathbf{u}_{n\nu}$ being the displacement of 
the $\nu$-th particle in the $n$-th unit cell of the lattice
from its equilibrium position. Finally, the
index $i$ denotes the Cartesian components,
$i=x,y$, and $\Phi({\bf r}^N)$ is the total potential energy of the crystal
at the harmonic approximation.
The dimensionless eigenvalues
$\lambda(\mathbf{q})\sigma^2/\varepsilon$ are determined for ${\bf q}$-values
along those axes of the first Brillouin zone that link the points of high
symmetry within the same.

In addition to affecting the ground states, 
the susceptibility ratio $m$ can be used to
fine-tune the appearance of a PBS corresponding to a given ground state, as
the stable configurations are robust against small variations in $m$. In the
following discussion, we focus on mixtures with $C\ge 1/2$, as the complexity
in the emerging patterns increases with the number of weak dipoles in the
system and the PBS's reflect recurrences in the ground-states. 

\begin{figure}
\includegraphics[width=8.6cm,clip]{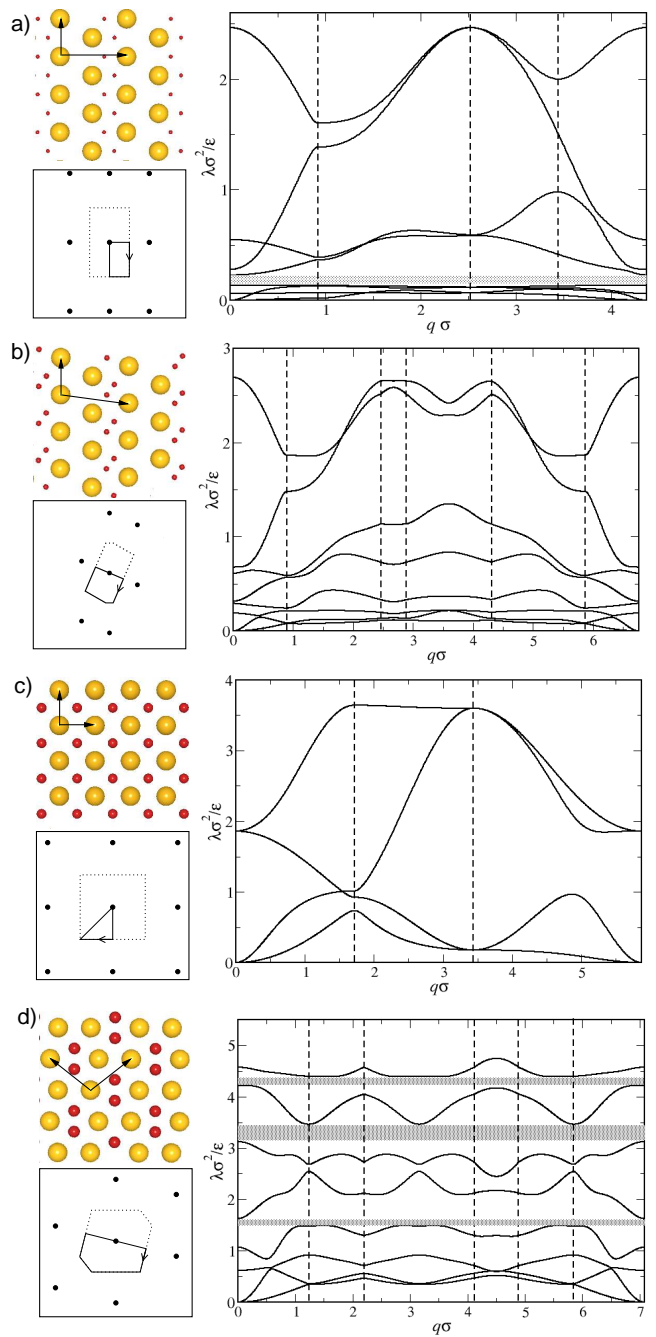}
\caption{Phonon band structures for a binary mixture of dipolar colloids with
  a concentration of $C=1/2$ and susceptibility ratios of (a) $m=0.018$, (b) $m=0.04$,
  (c) $m=0.18$, and (d) $m=0.28$. Band gaps are marked by shaded regions. The
  corresponding ground-state is depicted on the top left of each band
  structure, with the species of stronger (weaker) dipoles given by yellow
  (red) spheres and two lattice vectors marking the chosen unit cell. The
  sphere size reflects the dipole strength. On the bottom left of each band
  structure, a cell of the reciprocal lattice with its first Brillouin zone,
  marked by broken lines, is shown. The path of $\mathbf{q}$-values along
  which the band structures were determined, is indicated by a full
  line.\label{fig_C11_spectra}} \end{figure}

We start 
by demonstrating the influence of the susceptibility ratio $m$ on the
PBS for mixtures with a concentration of $C=1/2$. 
We vary the susceptibility ratio in a range from $m=0.003$ to
$m=0.41$ and calculate the PBS's for the ground states predicted 
in Refs.~\cite{For08LoV,For09}.
In Fig.~\ref{fig_C11_spectra} we show the phonon spectra
for $C=1/2$ and increasing susceptibility ratio from top
to bottom ($m=0.018$, $0.04$, $0.18$, and $0.41$). The corresponding
ground states, together with their first Brillouin zones and the path along
which the PBS's were determined, are depicted in Fig.~\ref{fig_C11_spectra}
as well.
At the chosen concentration, the dipolar mixture
develops three different types of ground states depending on $m$.
For very low $m$-values, the stronger dipoles
form a hexagonal lattice unaffected by the presence of the weaker species,
the latter arranging 
themselves in lanes and occupying interstitial sites. The associated PBS is
characterized by a nonuniform distribution of dispersion curves: the topmost
four branches span a much broader eigenvalue range than the remaining four
modes at the bottom of the band structure. In addition, the
`unfurled' upper branches are separated from the compressed lower region by a
distinctive gap, see Fig.~\ref{fig_C11_spectra}(a). Small increases in the
susceptibility ratio $m$ leave the ground state unchanged, despite of the fact that
interactions involving the weaker species intensify \cite{For08LoV,For09}. 
In the PBS, a small
increase in $m$ is reflected by a widening of the gap and a gradual expansion
of the lower dispersion curves. Above a certain threshold in $m$, the weak
dipoles are able to distort the hexagonal pattern of their stronger
counterparts. This decrease in symmetry in the ground state is also reflected
in the PBS, see Fig.~\ref{fig_C11_spectra}(b).
At susceptibility ratios between $m=0.06$ and $0.18$, a
regular square lattice is predicted to be the energetically most favorable
arrangement of dipoles \cite{For09}. 
In the PBS's associated to the quadratic configuration,
the four dispersion curves cover the eigenvalue range in a uniform
fashion: all branches are fully unfurled and no gaps occur. Variations in the
susceptibility ratio that leave the ground-state unchanged result in a
stretching of the PBS without altering its general appearance. 
As an example, Fig.~\ref{fig_C11_spectra}(c) shows the PBS of the square lattice formed at $m=0.18$.
For $m\ge0.28$, the ground-state of the binary mixture
changes from the regular square lattice to the $H_2$-structure \cite{Lik92},
which consists of distorted hexagons of strong dipoles accommodating two
particles of the weaker species. The PBS's of the $H_2$-lattice are again
characterized by the occurrence of distinct eigenvalue bands. The number of
these bands and the width of the gaps separating them from each other are 
tunable via the susceptibility ratio: at $m=0.28$, i.e., close to the
transition from the square to the $H_2$-lattice, three gaps are observed, see
Fig.~\ref{fig_C11_spectra}(d). 

\begin{figure}
\includegraphics[width=8.2cm,clip]{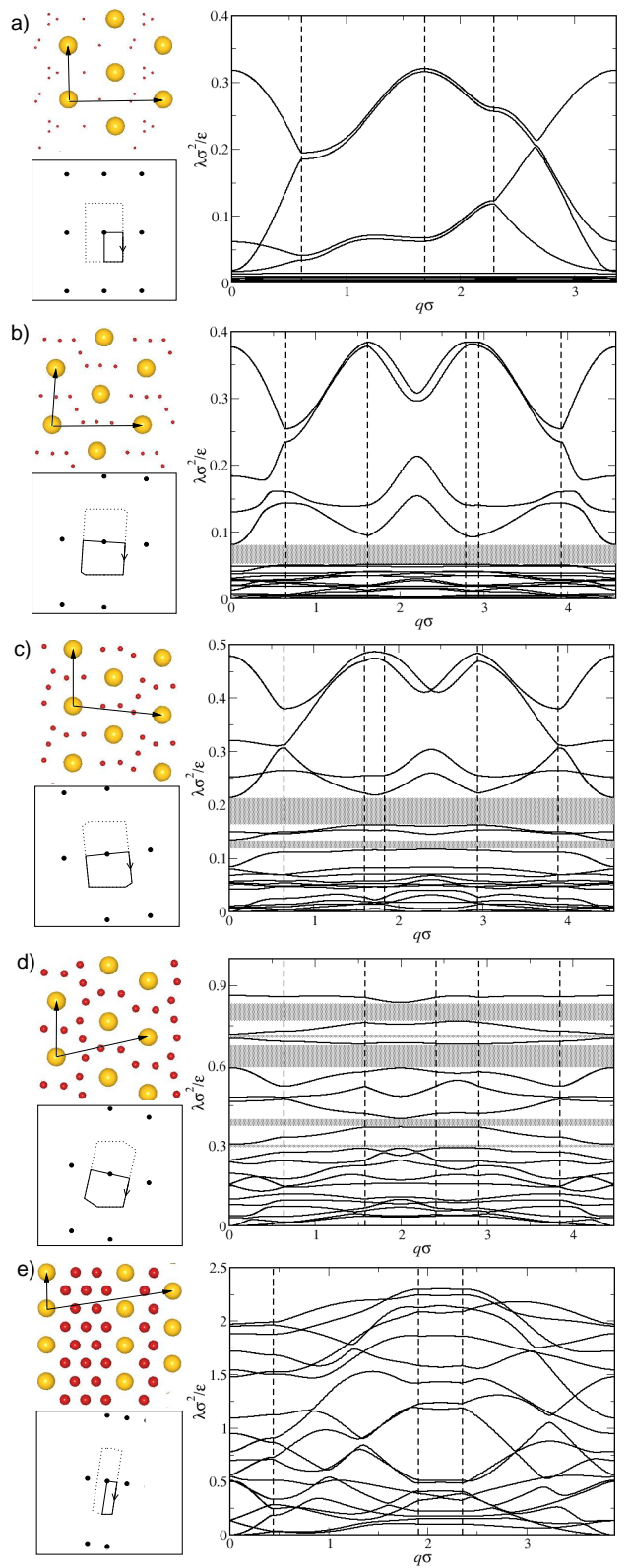}
\caption{Phonon band structures for a binary mixture of dipolar colloids with
  a concentration of $C=7/9$ and susceptibility ratios of (a) $m=0.003$, (b) $m=0.018$,
  (c) $m=0.04$, (d) $m=0.09$, and (e) $m=0.28$.\label{C27_spectra}}
\end{figure}

At higher concentrations of weak dipoles, more exotic particle arrangements
with a low degree of symmetry appear among the ground states. 
For the PBS's, the trends described above become more
pronounced. In Fig.~\ref{C27_spectra} we show selected PBS's obtained for mixtures
with $C=7/9$ and susceptibility ratios in the range from $m=0.003$ to
$m=0.28$. At small values of $m$, the non-uniform distribution of
dispersion curves is recovered. The four curves at the top of the PBS again cover a
much broader eigenvalue range than the rest of the branches and a gap
separating the compact from the extended dispersion curves is observed, even
at very low values of $m$. The susceptibility ratio can again be used to tune
the gap width, see Figs.~\ref{C27_spectra}(a), \ref{C27_spectra}(b). In contrast to
the previous case, $C=1/2$, the distortion of the underlying hexagonal lattice
induced by an increase of $m$, does not destroy the separation of the PBS into
distinct bands. A further increase in $m$ opens,
on the contrary, additional gaps, first, by
causing branches from the lower, compressed region to detach and then, as $m$
increases, by a progressive flattening of the topmost curves, 
going from Figs.~\ref{C27_spectra}(c) to \ref{C27_spectra}(d). 
At $m=0.09$, Fig.~\ref{C27_spectra}(d),
a maximum of five distinct bands is reached. If the susceptibility ratio
is increased above a certain threshold, i.e. $m\ge0.18$, the ground state of
the mixture undergoes a transition to lane-like arrangements, forcing all gaps
in the PBS to close and leading to a uniform distribution of well modulated
dispersion curves, Fig.~\ref{C27_spectra}(e).

\begin{figure}
\includegraphics[width=8.2cm,clip]{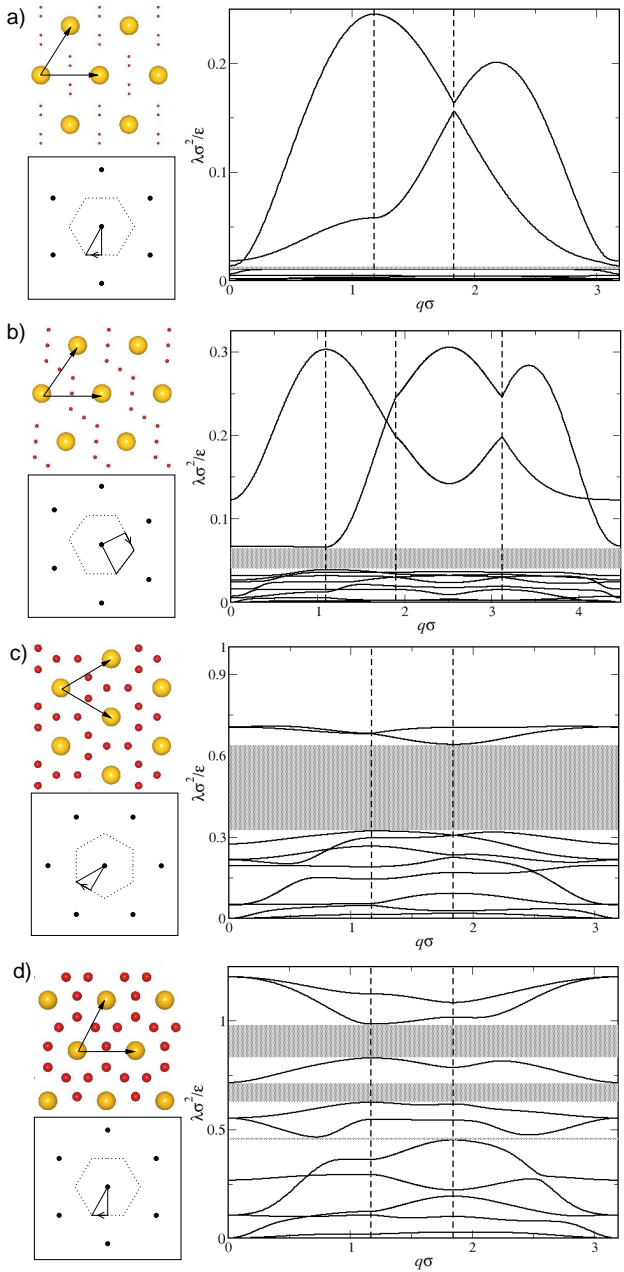}
\caption{Phonon band structures for a binary mixture of dipolar colloids with
  a concentration of $C=4/5$ and $m$-values: (a) $m=0.003$, (b) $m=0.018$,
  (c) $m=0.09$, (d) $m=0.18$.\label{C14_spectra}}
\end{figure}

The general trends become even clearer in the sequence of PBS's obtained at
$C=4/5$, see Fig.~\ref{C14_spectra}, for PBS calculated at $m=0.003$, $0.018$,
$0.09$, and $0.18$. The band structures at low values of $m$ are
characterized by two distinct bands, the width of the gap separating them is well
tunable via the susceptibility ratio, Figs.~\ref{C14_spectra}(a), \ref{C14_spectra}(b).
At intermediate values of $m$, the dispersion curves flatten out and
multiple gaps appear between the less modulated branches, see 
Figs.~\ref{C14_spectra}(c), \ref{C14_spectra}(d), which vanish again with
the onset of lane formation observed at high susceptibility ratios.  

Thus far, we have not considered a possible mass asymmetry of the two species in
our calculations. This is done because the colloidal systems that allow for an 
experimental realization of our findings are {\it Brownian systems}, in which
the mass of the particles is rendered irrelevant by overdamping \cite{Kei04,Gru07}. 
Including the
particle masses $M_\nu$ by multiplying expression (\ref{eqn_DynMatrix}) for the dynamical
matrix with $1/\sqrt{M_\nu M_{\nu'}}$, we can easily switch to {\it Newtonian
systems} in our theoretical approach and gain the 
mass ratio $M=M_B/M_A$ as an additional parameter to
tune the PBS's with. In ongoing investigations we have found that 
mass asymmetry (as it occurs, for instance, in dusty plasmas) 
enhances the effects described above: by lowering $M$, the topmost $2n_A$
branches can be decoupled from the rest of the band structure and shifted to
higher eigenvalues, thus opening a gap of variable width. In addition, the
flattening of dispersion curves gets more pronounced with increasing
mass-asymmetry, so that even 
completely flat branches can be induced
on the PBS's.

In summary, we have investigated the possibility to control the PBS's of a
dipolar binary mixture via the parameters determining the interactions and the
ground-state of the system. By calculation of the dispersion curves for the
stable particle arrangements, we could show that, for systems with 
a majority of weaker dipoles,
multiple gaps can be opened by suitable adjustments in the susceptibility
ratio $m$. In addition, remarkably flat branches were observed in the PBS
featuring multiple gaps. Our results are easily verifiable in experimental setups
that employ two-dimensional binary magnetic colloids, where dispersion
curves can be measured by means of the equipartition theorem~\cite{Kei04,Gru07}.

This work has been supported by the FWF under Proj.~No.~P17823
and by the SFB-TR6, Project C3.
\bibliographystyle{prsty} 


\end{document}